\documentclass[twocolumn]{aastex63}
\pdfoutput=1
\usepackage{graphicx, url}
\usepackage{enumerate, enumitem}
\usepackage{amssymb, amsmath, amsfonts, upgreek}
\usepackage{gensymb}
\usepackage{mathtools}
\usepackage{hyperref}
\usepackage{xspace}
\usepackage{xcolor}
\usepackage{ulem}

\newcommand{\masyr}{\hbox{mas\,yr$^{-1}$}}

\newcommand{\Lsun}{\mbox{$L_{\sun}$}}
\newcommand{\Lbol}{\mbox{$L_{\rm bol}$}}
\newcommand{\Msun}{\mbox{$M_{\sun}$}}

\newcommand{\Mjup}{\mbox{$M_{\rm Jup}$}}

\newcommand{\pmoffs}[2]{^{+ #1}_{- #2}}

\newcommand{\Hipparcos}{\textsl{Hipparcos}\xspace}
\newcommand{\hipparcos}{\textsl{Hipparcos}\xspace}
\newcommand{\Gaia}{\textsl{Gaia}\xspace}
\newcommand{\gaia}{\textsl{Gaia}\xspace}

\newcommand{\htofcodename}{\texttt{htof}\xspace}
\newcommand{\fiducialmass}{$9.6 \pmoffs{1.9}{1.8} \, \Mjup$\xspace}
\newcommand{\worstcasemass}{$9.4 \pmoffs{2.2}{2.1} \, \Mjup$\xspace}
\newcommand{\htofversion}{\href{https://github.com/gmbrandt/HTOF/tree/0.3.4}{0.3.4}\xspace}

\received{April 7, 2021}
\revised{May 14, 2021}
\accepted{May 26, 2021}

\shorttitle{The First Dynamical Mass Measurement in the HR 8799 System}
\shortauthors{Brandt et al.}
\submitjournal{ApJ Letters}

\begin{document}

\title{The First Dynamical Mass Measurement in the HR 8799 System}

\author[0000-0003-0168-3010]{G.~Mirek Brandt}
\altaffiliation{NSF Graduate Research Fellow}
\affiliation{Department of Physics, University of California, Santa Barbara, Santa Barbara, CA 93106, USA}

\author[0000-0003-2630-8073]{Timothy D.~Brandt}
\affiliation{Department of Physics, University of California, Santa Barbara, Santa Barbara, CA 93106, USA}

\author[0000-0001-9823-1445]{Trent J.~Dupuy}
\affiliation{Institute for Astronomy, University of Edinburgh, Royal Observatory, Blackford Hill, Edinburgh, EH9 3HJ, UK}

\author[0000-0002-7618-6556]{Daniel Michalik}
\affiliation{European Space Agency (ESA), European Space Research and Technology Centre (ESTEC), Keplerlaan 1, 2201 AZ Noordwijk, The Netherlands}

\author[0000-0002-2919-7500]{Gabriel-Dominique Marleau}
\affiliation{%
Institut f\"ur Astronomie und Astrophysik,
Universit\"at T\"ubingen,
Auf der Morgenstelle 10,
72076 T\"ubingen, Germany
}
\affiliation{%
Physikalisches Institut,
Universit\"{a}t Bern,
Gesellschaftsstr.~6,
3012 Bern, Switzerland
}
\affiliation{%
Max-Planck-Institut f\"ur Astronomie,
K\"onigstuhl 17,
69117 Heidelberg, Germany
}

\begin{abstract}
HR 8799 hosts four directly imaged giant planets, but none has a mass measured from first principles. 
We present the first dynamical mass measurement in this planetary system, finding that the innermost planet HR~8799~e has a mass of \fiducialmass. This mass results from combining the well-characterized orbits of all four planets with a new astrometric acceleration detection (5$\sigma$) from the \gaia EDR3 version of the \hipparcos-\gaia Catalog of Accelerations. We find with 95\% confidence that HR~8799~e is below $13\, \Mjup$, the deuterium-fusing mass limit. 
We derive a hot-start cooling age of $42\pmoffs{24}{16}$\,Myr for HR~8799~e that agrees well with its hypothesized membership in the Columba association but is also consistent with an alternative suggested membership in the $\beta$~Pictoris moving group.
We exclude the presence of any additional $\gtrsim$5-$\Mjup$ planets interior to HR~8799~e with semi-major axes between $\approx$3--16\,au. We provide proper motion anomalies and a matrix equation to solve for the mass of any of the planets of HR~8799 using only mass ratios between the planets.
\end{abstract}


\section{Introduction}
HR 8799 is the only star that has been detected with four directly imaged exoplanets \citep{Marois1348, 2010Natur.468.1080M, 2011ApJ...729..128C}. Each planet orbits counter-clockwise on the sky, in a gently eccentric orbit, with a nearly face-on configuration \citep{2012ApJSudol_hr8799_stability1, 2015ApJPueyo_etal, 2016_Konopacky_etal, 2016arXiv160703980C}. 
The orbits of all four planets are known with exquisite precision. \cite{2018AJ....156..192W} studied the system in detail and restricted orbital parameter space based on configurations that are dynamically stable. There have been many mass constraints based on the long term dynamical stability of the system (e.g., all planets $\lesssim5\Mjup$ by \citealp{2013A&AEsposito_etal}; inner three planets $< 13 \Mjup$ by \citealp{2015ApJPueyo_etal}). However, there are no robust dynamical, i.e., from Newtonian mechanics, mass measurements for any of the planets in the HR~8799 system.

The HR~8799 planets induce accelerations on their host star that are measurable by sufficiently precise absolute astrometry, enabling direct, dynamical measurements of the planets' masses. The \hipparcos-\gaia catalog of accelerations \citep[HGCA,][]{brandt_cross_cal_gaia_2018, 2021arXiv_HGCAEDR3} has cross-calibrated the positions and proper motions so that these accelerations can be used for inference. The acceleration of HR~8799~A was not significant in the \gaia DR2 \citep{Gaia_Astrometry_2018} version of the catalog \citep{brandt_cross_cal_gaia_2018}. However, \Gaia's precision on bright stars has increased significantly in EDR3 (\citealp{2020GaiaEDR3_catalog_summary, Lindegren+Klioner+Hernandez+etal_2020, 2021arXiv_HGCAEDR3}). In the \gaia EDR3 version of the HGCA, HR~8799~A is accelerating at nearly 5$\sigma$. In this paper we infer the mass, and thereafter the age, of HR~8799~e from this acceleration.

The HR~8799 system is most commonly thought to be an $\approx40$-Myr-old member of the Columba association \citep{2011ApJ...732...61Z}, but \citet{2019MNRAS.489.2189L} recently suggested membership with the younger $\beta$~Pic Moving Group (BPMG). Given a $\approx$40-Myr age, hot-start evolutionary models predict masses of $7\pm2$\,\Mjup\ for the innermost three planets, well below the deuterium-fusion mass boundary \citep[$\approx$12-13\,\Mjup;][]{2011ApJ...727...57S}. Lower masses ($\lesssim$5--7 \Mjup) improve the system's dynamical stability \citep{Fabrycky_2010, 2012ApJSudol_hr8799_stability1}, but resonant locking could render the system stable at higher planet masses \citep{2016A&A...592A.147G, 2018AJ....156..192W, Gozdziewski+Migaszewki_2018,Gozdziewski+Migaszewski_2020}. Higher masses would imply either an older age for the system or entropy loss during formation (a colder start), although there is a limit to the entropy that can be lost at formation \citep{Marleau+Cumming_2014}. Conversely, lower masses would suggest a younger age and therefore favor membership with the BPMG. Dynamical mass measurements can conclusively test such hypotheses. The inferred atmospheric properties and chemical abundances of the planets also depend on their assumed masses (e.g., \citealp{2020AJ.Jwang.ChemicalAbundance, 2020A&A...640A.131M}).

We detail our method in Section~\ref{sec:methodology}. In Section~\ref{sec:results} we consider planet mass ratios estimated from their relative luminosities, show that the mass of HR~8799~e is insensitive to these mass ratios, and obtain a robust dynamical mass measurement. We derive a substellar cooling age for HR~8799~e in Section~\ref{sec:age} and conclude in Section~\ref{sec:conclusions}.

\section{Methodology}\label{sec:methodology}

The acceleration that HR~8799~A experiences is the sum of the acceleration due to each of its four planetary companions.  Because the planet masses are small compared to the mass of HR~8799~A ($1.47\pmoffs{0.11}{0.08}$\,\Msun; \citealp{2018AJ....156..192W}), the star's motion is approximately given by a linear combination of the orbits of the planets, weighted by their masses.  We can then optimize the planet masses until the modelled host star acceleration matches the observed value.

We use all 1000 samples from the orbital posteriors from \cite{2018AJ....156..192W}, published on the github repository for the resource \url{whereistheplanet.com} \citep{2021ascl.soft01003W}. The orbits correspond to the coplanar, dynamically stable case of the HR~8799 system \citep[see Table 4 of][]{2018AJ....156..192W}. Each set of orbital parameters, together with the masses of the four planets and the star, predicts the motion of HR~8799~A.  

\cite{2018AJ....156..192W} used the \gaia DR1 parallax of $24.76 \pm 0.64$~mas \citep{Gaia_General_2016}\footnote{In all cases where we quote a posterior by listing $m\pmoffs{u}{l}$ or $m \pm \sigma$: $m$ denotes the median with $l$ and $u$ (or singularly, $\sigma$) denoting the 16\% and 84\% confidence intervals, respectively.}. \gaia EDR3 measures a much more precise value of $24.462 \pm 0.046$~mas \citep{Lindegren+Klioner+Hernandez+etal_2020}. We scale all 1000 MCMC samples to the \gaia EDR3 parallax, removing its contribution to the uncertainty of the orbital fits. Defining $r$ to be the ratio of a given chain's parallax to the \gaia EDR3 value, we multiply the semimajor axes by $r$ (to preserve the relative astrometry) and multiply the system mass by $r^3$ (to preserve the orbital periods). We keep all other orbital parameters unchanged. Updating the parallax to its \gaia EDR3 value ultimately improves the fractional error on the final mass estimate from 20\% to 19.5\%.

We use the open-source tool \htofcodename \citep{MirekHTOFtemporary, htof_zenodo, 2021arXiv_orvara} to model \hipparcos and \gaia observations. In brief, \htofcodename uses the \hipparcos intermediate astrometric data (both \cite{HIP_TYCHO_ESA_1997} and \cite{vanLeeuwen_2007} reductions) and predicted scan angles and observational epochs of \gaia (via GOST\footnote{\url{https://gaia.esac.esa.int/gost/}}), to generate synthetic \hipparcos and \gaia astrometry for any orbit.

The astrometric measurement that we use is the proper motion anomaly. This is the difference between a nearly instantaneous proper motion from \gaia EDR3 and a long-term proper motion (the difference in position between the \hipparcos and \gaia astrometry missions divided by the time between the measurements).  We denote these proper motion anomalies as, e.g.,
\begin{equation}\label{eq:pmra_anomaly}
    \Delta \mu_{\alpha*} = \mu_{\it Gaia} - \frac{{\alpha*}_{\it Gaia} - {\alpha*}_{\it Hip}}{t_{\it Gaia} - t_{\it Hip}}
\end{equation}
where $\alpha* = \alpha \cos \delta$. Parameters with subscript ${\it Hip}$ refers to the average of those parameters from \hipparcos 2007 and 1997, weighted 60/40 as they are in the HGCA \citep{brandt_cross_cal_gaia_2018}. A set of orbital parameters then gives predicted values for $\Delta \mu_{\alpha*}$ and $\Delta \mu_{\delta}$ as functions of the masses of the planets, $m_b, m_c, m_d$, and $m_e$.  Because HR~8799~A's motion closely follows a linear combination of Keplerian orbits, we can represent its predicted proper motion anomaly as the Jacobian 
\begin{align}
\begin{bmatrix}
    \Delta \mu_{\alpha*} \\
    \Delta \mu_{\delta}
\end{bmatrix}_{\rm model}
=
\begin{bmatrix}
    \frac{\partial \Delta \mu_{\alpha*}}{\partial m_b} & 
    \frac{\partial \Delta \mu_{\alpha*}}{\partial m_c} & 
    \frac{\partial \Delta \mu_{\alpha*}}{\partial m_d} & 
    \frac{\partial \Delta \mu_{\alpha*}}{\partial m_e} \\
    \frac{\partial \Delta \mu_{\delta}}{\partial m_b} & 
    \frac{\partial \Delta \mu_{\delta}}{\partial m_c} & 
    \frac{\partial \Delta \mu_{\delta}}{\partial m_d} & 
    \frac{\partial \Delta \mu_{\delta}}{\partial m_e}
\end{bmatrix}
\begin{bmatrix}
    m_b \\
    m_c \\
    m_d \\
    m_e
\end{bmatrix}.
\label{eq:jacobian}
\end{align}
We compute the partial derivatives by using only the orbit of a given planet and assigning that planet unit mass. We do this for all 1000 orbital draws. We use {\tt REBOUND} and the {\tt ias15} scheme \citep{rebound_2012_main, rebound_ias15} to integrate the orbits in time.

When we compute the model partial derivatives, we mix the \cite{HIP_TYCHO_ESA_1997} and \cite{vanLeeuwen_2007} positions according to the same 60/40 ratio adopted by the HGCA. \citet{brandt_cross_cal_gaia_2018} show in their Section 7 and Figure 2 that a 60/40 mix of the two \Hipparcos reductions' proper motion measurements better matches the long-term proper motions between \Hipparcos and \Gaia than either reduction on its own. The EDR3 version of the HGCA \citep{2021arXiv_HGCAEDR3} confirms this finding and also shows that a 60/40 mix of the two \Hipparcos reductions' position measurements best matches the \Gaia positions extrapolated back to the \Hipparcos observational epoch. We compute our positions at the same central epochs as given in the HGCA\footnote{These central epochs are, in years for dec. and right-ascension: 2015.85 2015.76 for \gaia EDR3 and 1991.35 1991.34 for \hipparcos.}. This forward modeling allows us to directly compare our proper motion anomalies to the values given in the HGCA. The HGCA is calibrated so that the measured proper motion anomalies have Gaussian uncertainties. We can therefore identify $\chi^2$ with $-2 \ln {\cal L}$ and find the masses by maximizing the likelihood ${\cal L}$, or minimizing

\begin{equation}
\chi^2 = -2 \ln {\cal L} = 
{\bf d}^T
\left({\bf C}_{HG} + {\bf C}_{\it Gaia} \right)^{-1}
{\bf d}
\label{eq:chisq}
\end{equation}
where ${\bf C}_{HG}$ is the HGCA covariance matrix for the two \hipparcos-\gaia long-term proper motions, ${\bf C}_{\it Gaia}$ is the HGCA covariance matrix for the two \gaia EDR3 proper motions, and
\begin{equation}
    {\bf d} = \begin{bmatrix}
    \Delta \mu_{\alpha* {\rm model}} - \Delta \mu_{\alpha*  {\rm HGCA}} \\
    \Delta \mu_{\delta  {\rm model}} - \Delta \mu_{\delta  {\rm HGCA}}
\end{bmatrix}.
\label{eq:resid}
\end{equation}
In Equation \eqref{eq:resid}, $\Delta \mu_{\alpha*  {\rm model}}$ and $\Delta \mu_{\alpha*  {\rm model}}$ are calculated from the right-hand side of Equation \ref{eq:jacobian}. ${\bf C}_{HG}$ and ${\bf C}_{\it Gaia}$ are given in the HGCA. We republish their sum here, along with the anomalies, for ease of reproducibility:
\begin{align}
    {\bf C}_{HG} + {\bf C}_{\it Gaia} &= \begin{bmatrix}
    6.5934 & 0.7473 \\
    0.7473 & 6.9888
\end{bmatrix} 10^{-3} \left({\rm mas\,yr}^{-1}\right)^2 \\ \begin{bmatrix}
    \Delta \mu_{\alpha*  {\rm HGCA}} \\
    \Delta \mu_{\delta  {\rm HGCA}}
\end{bmatrix} &= \begin{bmatrix}
    -0.268  \\
    -0.348
\end{bmatrix} \masyr . \label{eq:pmanomaly_HGCA}
\end{align}
We compute the partial derivatives of Equation \eqref{eq:chisq} against $\Delta \mu_{\alpha*}$, and $\Delta \mu_{\delta}$. Setting these partials to zero gives an under-constrained system of equations. The two components of the proper motion anomaly measure a combination of the masses of the four planets, but not the four masses individually. However, if we assume a relationship between the individual masses, then Equation \eqref{eq:chisq} produces an over-constrained system for the masses. If we take a uniform prior on one planet mass and assume mass ratios for the remaining three planets, Equation \eqref{eq:chisq} represents a Gaussian mass posterior for each set of orbital parameters.  

\section{Results}\label{sec:results}
In this section we use Equations \eqref{eq:jacobian}--\eqref{eq:pmanomaly_HGCA}, together with varying assumptions about the mass ratios of the four planets, to derive constraints on the mass of HR~8799~e.

\subsection{Fixed mass ratios}

We initially assume the fixed mass ratios derived by  \cite{2018AJ....156..192W} from the observed luminosities assuming hot-start models,
\begin{equation}
    m_e = m_d = m_c = 1.25 m_b.
    \label{eq:massratios_fiducial}
\end{equation}

With the assumed ratios of the masses of the four planets, Equation \eqref{eq:chisq} fits one free parameter to two covariant data points. We begin with the calculation of the Jacobians, Equation \eqref{eq:jacobian} (one for each of the 1000 orbital draws).  We fit 1000 Jacobians and sum the posteriors. The resulting posterior is very nearly Gaussian with mean $9.59 \Mjup$ and standard deviation $1.84 \Mjup$. The residual from the best-fit mass should be $\chi^2$-distributed with one degree of freedom.  The best-fit $\chi^2$ is only 0.05: the observed astrometric acceleration from the HGCA agrees almost perfectly with the model prediction using the \cite{2018AJ....156..192W} orbits and mass ratios.

Because the orbital elements of the HR~8799 planets are so well characterized, the derivatives of the anomalies with respect to each planet's mass vary little between the  MCMC draws. The element-wise median Jacobian matrix is an approximation to all 1000 sets of orbital parameters; it is given by

\begin{equation}\label{eq:median_Jacobian}
\frac{{\bf J}_{\rm median}}{\rm \upmu as\,yr^{-1}\,M_{\rm Jup}^{-1}}
=
\begin{bmatrix*}
    2.00 & 
    -3.94 & 
    \hspace{1.8mm}-3.01 & 
    -24.20 \\
    1.14 & 
    \phantom{-} 3.80 & 
    -13.74 & 
    -26.05
\end{bmatrix*}.
\end{equation}

The width of our posterior is dominated by the observational uncertainty in absolute astrometry from the HGCA, which corresponds to about $\pm 1.8 \,\Mjup$ (20\%). The uncertainty in the mass of HR~8799~e from the orbital draws alone is just $0.35\,\Mjup$ (enlarging the final error bars by only 2\%). Using the median Jacobian (Equation \eqref{eq:median_Jacobian}) together with the mass ratios in Equation \eqref{eq:massratios_fiducial}, and broadening with the orbital motion uncertainty ($0.35\,\Mjup$), gives a posterior that is indistinguishable from the full posterior using 1000 sets of orbital parameters.

We use the median Jacobian matrix for the remainder of this work and convolve our posterior mass distributions for HR~8799~e with a Gaussian of standard deviation $0.35\,\Mjup$ (which accounts for the negligible contribution from orbital motion uncertainty).

\subsection{Varying mass ratios}
The mass ratios of the four planets are known only to roughly $\pm 15\%$ from hot-start models \citep{2018AJ....156..192W}. We now show that the mass of HR~8799~e is robust to larger variations in the mass ratios.
We quantify changes in the mass ratios by three coefficients $\gamma_i$, which we define by

\begin{equation}
    m_e = \gamma_d m_d = \gamma_c m_c = \gamma_b 1.25 m_b.
    \label{eq:massratios_final}
\end{equation}
Setting $\gamma_d = \gamma_c = \gamma_b = 1$ yields the fiducial mass ratios.

\begin{figure}
    \centering
    \includegraphics[width=\linewidth]{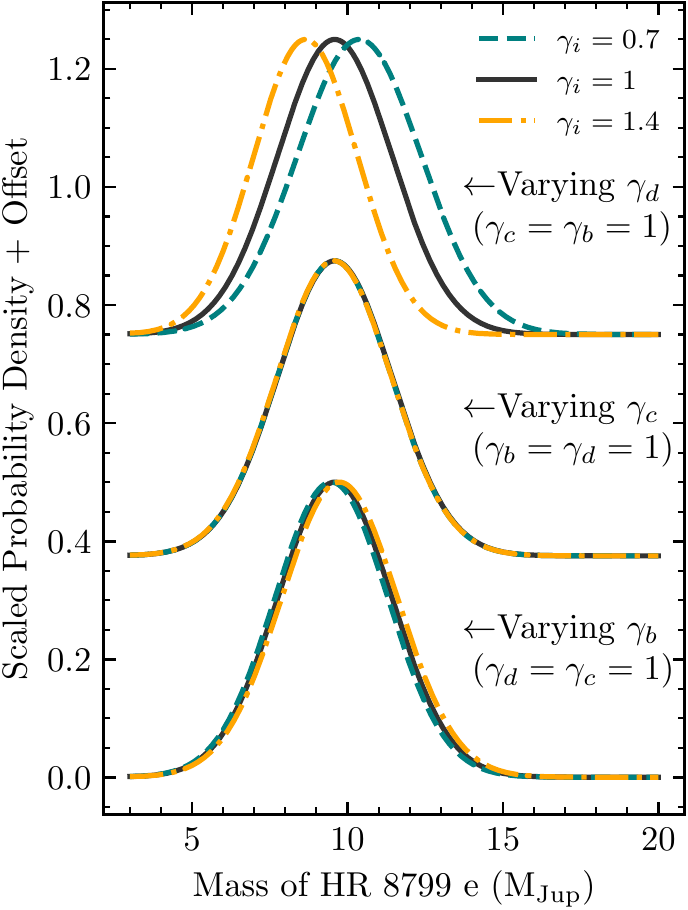}
    \caption{
    Posteriors for the mass of HR~8799~e, varying the assumed mass ratio of each other planet relative to planet e from 70\% ($\gamma = 0.7$, teal dashed lines) to 140\% ($\gamma = 1.4$, orange dot-dashed lines) of its fiducial value (see Equation \eqref{eq:massratios_final}).  The mass of HR~8799~e is moderately covariant with that of planet d but insensitive to the masses of planets b and c.}
    \label{fig:mass_pdf_comparing_ratios}
\end{figure}

Figure~\ref{fig:mass_pdf_comparing_ratios} shows the mass posterior of HR~8799~e assuming nine different mass ratios. We vary each of the $\gamma_i$ independently between 0.7 to 1.4 (corresponding to varying the masses of b,c, and d by roughly $\pm 4\,\Mjup$).
Varying the mass ratio of planet b or c to planet e by this amount changes the inferred mass of HR~8799~e by $\lesssim$0.1\,$\Mjup$, or $\lesssim$0.05$\sigma$. The top curves in Figure~\ref{fig:mass_pdf_comparing_ratios} show that the mass of HR~8799~e is covariant with HR~8799~d but that the variation is $\approx$1\,$\Mjup$, or $\approx$0.5$\sigma$, even with the extreme range of mass ratios presented.
Taylor expanding the maximum likelihood mass, $\langle m_e \rangle$, about our base case mass ratios $\gamma_d = \gamma_c = \gamma_b = 1$ yields

\begin{align}\label{eq:empirical_mass_of_e}
    \frac{\langle m_e \rangle}{\Mjup} &\approx 9.56 - 2.50 \left(\gamma_d - 1 \right) \nonumber \\ &- 0.001 \left(\gamma_c - 1 \right) + 0.37\left(\gamma_b - 1 \right).
\end{align}

The mass of HR~8799~d is moderately covariant with that of HR~8799~e, while HR~8799~b and HR~8799~c have little effect: b is too far away, while c induces a proper motion anomaly nearly perpendicular to that induced by e.

We now generalize Figure~\ref{fig:mass_pdf_comparing_ratios} by marginalizing over the possible range of mass ratios (i.e., effectively summing the posteriors of Figure~\ref{fig:mass_pdf_comparing_ratios}). We use independent, log-normal priors (base-$e$ lognormal) centered on unity for each of $\gamma_d$, $\gamma_c$, and $\gamma_b$.  For our fiducial case, we take the $\gamma_i$ priors to have standard deviation 0.15 of the natural logarithm (0.065 dex). This corresponds to $\pm 15\%$, reflecting hot-start uncertainties \citep{2018AJ....156..192W}. We also include a worst-case where we use a logarithmic prior with a standard deviation of 0.45 (0.2 dex). This allows for deviations in the mass ratios of roughly $-35\%$ and $+55\%$, slightly more than that shown in Figure~\ref{fig:mass_pdf_comparing_ratios}. 

\begin{figure}
    \centering
    \includegraphics[width=\linewidth]{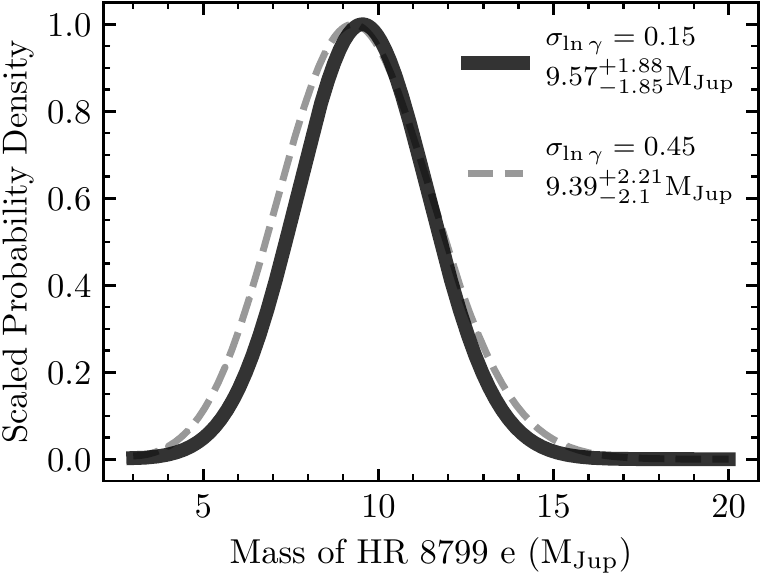}
    \caption{The mass posterior of HR~8799~e after marginalizing over bcd-to-e mass ratios (quantified by three $\gamma_i$ coefficients), each with a lognormal prior. Each posterior is labelled according to both its mass-ratio prior and the resulting posterior (non-rounded for convenience). $\sigma_{\ln \gamma}$ is the natural logarithmic standard deviation of the log-normal prior on the $\gamma_i$. Each prior has a natural logarithmic mean of 0.}
    \label{fig:mass_pdf_marginalized_mass_ratios}
\end{figure}

Figure~\ref{fig:mass_pdf_marginalized_mass_ratios} shows the posteriors on planet e's mass under the two prior choices. Our preferred result is a nearly Gaussian posterior of \fiducialmass. Adopting our worst-case prior, allowing for three times the range of mass ratios, yields \worstcasemass.

\subsection{Additional companions}

\begin{figure}
    \centering
    \includegraphics[width=\linewidth]{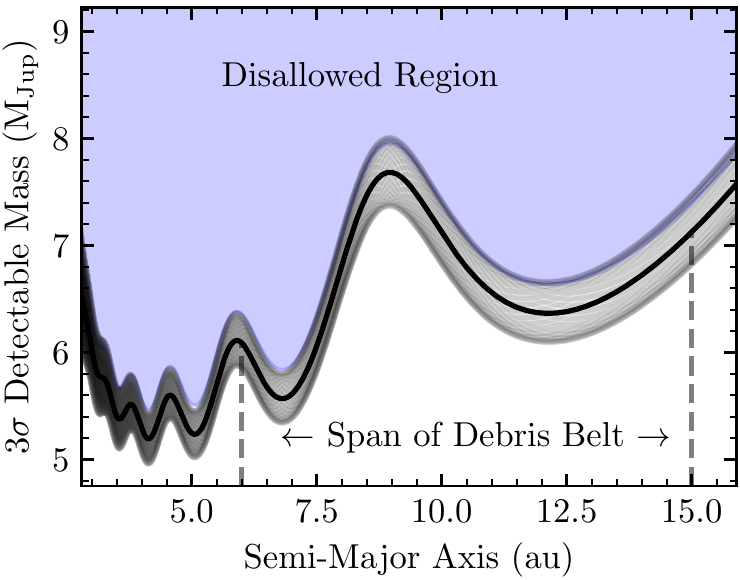}
    \caption{The minimum masses and semi-major axes of additional, unseen HR~8799 companions that would have been detected at $3\sigma$ using our HGCA \gaia EDR3-\hipparcos proper motion anomalies. The individual grey lines (forming a grey band together) show the 3$\sigma$ limits assuming a range of argument of periastron ($\omega$) from 0 to 2$\pi$. The black line is the 3$\sigma$ limit averaged over the all possible $\omega$.
    Planets lying in the blue ``disallowed region'' are excluded with at least 99.7\% confidence, regardless of their orbital phase. The approximate range of the inner debris belt is indicated by vertical dashed lines (6 to 15 au; \citealp{2020A&A...638A..50F}).}
    \label{fig:detectability_plot}
\end{figure}

An additional companion is detectable if it causes a significant astrometric perturbation on the host star. Figure~\ref{fig:detectability_plot} shows the semi-major axes and masses that an additional, unseen massive companion would need to cause perturbation large enough to be detected in \gaia EDR3. Planets in the blue region above the grey band would have yielded a significant (3$\sigma$) astrometric acceleration on the system that we would have seen in our analysis. Additional planets in the parameter space below the grey band (white region) are not excluded. We conclude that additional, unseen massive planets orbiting between 3 and 8\,au with masses exceeding 6\,$\Mjup$ are unlikely, as well as $\gtrsim 7\,\Mjup$ companions between 8\,au and the orbit of HR~8799~e ($\approx$16\,au). We therefore exclude the presence of any $\gtrsim$7\,$\Mjup$ companions orbiting amidst the inner debris belt, which spans 6--15\,au \citep{2020A&A...638A..50F}.
Our detection limits complement the findings by \citet{2021Wahhaj_HR8799_starhopping}, who excluded the presence of hot-start planets more massive than $\gtrsim 3\,\Mjup$ at two separations: 7.5 and 9.7 au.

\begin{figure}
    \centering
    \vskip -0.05in
    \includegraphics[width=0.95\linewidth]{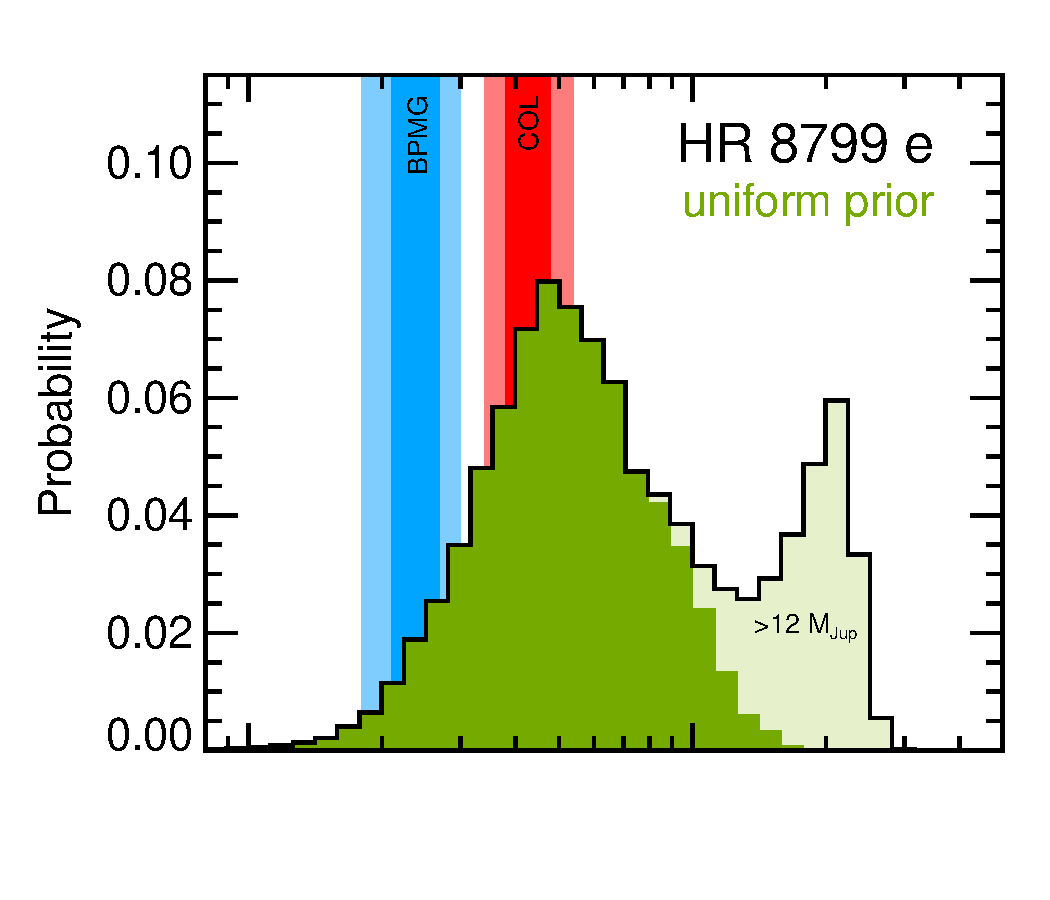}
    \vskip -0.3in
    \includegraphics[width=0.95\linewidth]{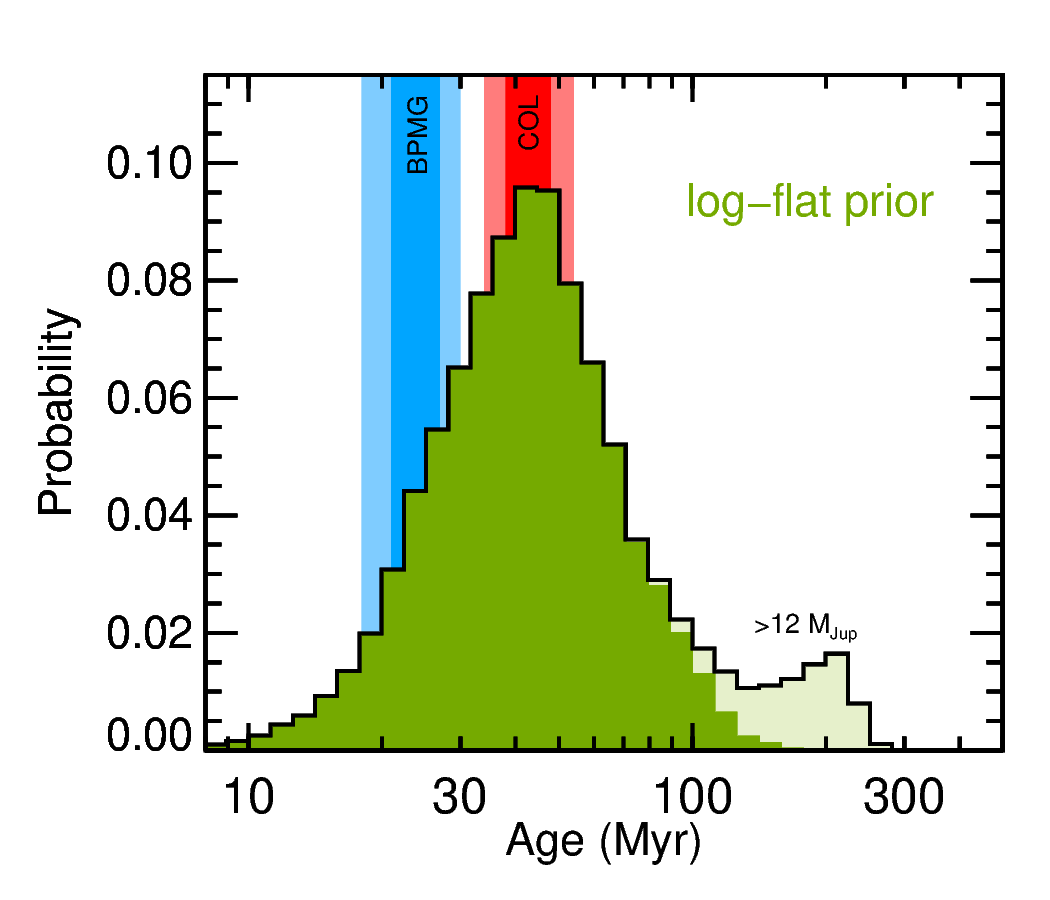}
    \vskip -0.15in
    \caption{Substellar cooling age for HR~8799~e derived from hot-start \citet{Saumon+Marley_2008} hybrid models using its observed luminosity and dynamical mass, with either a uniform (top panel) or log-flat (bottom panel) prior on age. 
    The 1$\sigma$ and 2$\sigma$ age ranges for the Columba association (red) and the BPMG (blue) are displayed for comparison, and our cooling age is consistent with both. 
    The peak at $\approx$200\,Myr corresponds to the 10\% of our mass posterior above 12\,\Mjup\ (lighter shading). The maximum likelihood age is 40--50\,Myr regardless of the age prior.}
    \label{fig:age}
\end{figure}

\section{The Age of the HR~8799 System}\label{sec:age}

With our dynamical mass for HR~8799~e, we infer the first cooling age for the planet and thereby the system. Prior to \citet{2011ApJ...732...61Z} identifying it as a member of the Columba association ($42\pmoffs{6}{4}$\,Myr; \citealt[]{Bell+Mamajek+Naylor_2015}), its age was only loosely constrained \citep{Marois1348}. 
\citet{2019MNRAS.489.2189L} recently suggested that it is actually a member of the younger $\beta$~Pictoris moving group \citep[BPMG, $24\pm3$\,Myr;][]{Bell+Mamajek+Naylor_2015}. 
Moreover, HR~8799 was one of four (out of 23) Columba members that \citet{2018ApJ...856...23G} chose not to use in their BANYAN~$\Sigma$ model due to being outliers, despite it still being considered a bona fide member. 

We perform a rejection-sampling analysis using mass and \Lbol, in a similar fashion as \citet{Dupuy+Liu_2017} and Brandt et al.\ (2021, submitted), to derive a hot-start cooling age for HR~8799~e. The HR~8799 planets are too luminous to be consistent with the very low initial entropies predicted by the \citet{Marley+Fortney+Hubickyj_2007} cold-start models \citep{Marleau+Cumming_2014}. Warm- and hot-start scenarios are allowed by the data, and simulations from \citet{berardo17}, \citet{berardocumming17}, and \citet{marleau19b} tend to favor the hot-start scenario in general.

We randomly draw masses from our posterior distribution and ages distributed uniformly or log-flat, then bi-linearly interpolate the evolutionary model grid and compute a test \Lbol. We accept or reject trials in a Monte Carlo fashion depending on how well the trials agree with the observed \Lbol.

We derive a new \Lbol\ for HR~8799~e using the absolute magnitude--\Lbol\ relations of \citet{Dupuy+Liu_2017}, the SPHERE photometry from \citet{2021Wahhaj_HR8799_starhopping}, and the $K$-band spectrum from \citet{2019A&A...623L..11G}. Although the relations of \citet{Dupuy+Liu_2017} are derived from field dwarfs, Figure~12 of \cite{2015ApJ...810..158F} demonstrates that young and field objects share the same $K$-band bolometric corrections within the 0.25\,mag scatter of their relations; we adopt this scatter as our uncertainty. We find $K_{\rm MKO} = K_{\rm S, 2MASS} = 16.00\pm0.02$\,mag and $\log(\Lbol/\Lsun) = -4.52\pm0.10$\,dex, which is consistent with but twice as precise as the measurement by \citet{2010Natur.468.1080M}.

Figure~\ref{fig:age} shows posterior distributions of the system age for two different age prior choices (uniform and log-flat). Both posteriors peak at 40--50\,Myr but differ at the young and old extremes. The choice of prior significantly affects the old extreme of the posterior. However, under either prior, Columba's age agrees well and BPMG's is consistent (1.2--1.7$\sigma$). 

Substellar cooling alone does not preclude older ages ($\gtrsim$100\,Myr), which have been shown to yield unstable orbits at the correspondingly higher masses ($>12$\,\Mjup). 
The high end of our mass posterior yields a smaller age peak at $\approx$200\,Myr that corresponds to a resurgence in luminosity at older ages due to deuterium fusion. Even though 9.8\% of our dynamical mass posterior for HR~8799~e lies above $12$\,\Mjup\, (where deuterium burning is possible), the planets of the HR8799 system have not been considered deuterium-fusing objects, with masses below $\approx$13\,\Mjup, based on their luminosity and hypothesized youth \citep{Marois1348, 2010Natur.468.1080M}. As well, they are unlikely to have masses in excess of $13$\,\Mjup\, on the basis of stability \citep{2015ApJPueyo_etal, 2018AJ....156..192W}.

Excluding masses above $12$\,\Mjup\, yields an age distribution that is approximately Gaussian in $\log{t}$ for both priors (see dark-shaded posteriors in Figure~\ref{fig:age}). The resulting age posterior (under the log-flat prior) is $\log(t/{\rm yr}) = 7.62\pm0.20$\,dex ($42\pmoffs{24}{16}$\,Myr).

If the HR~8799 system is indeed a BPMG member, it would be coeval and perhaps co-compositional with the giant planets $\beta$~Pic~b and c that have dynamical masses of $9.3\pmoffs{2.6}{2.5}$\,\Mjup\, and $8.3\pm1.0$\,\Mjup, respectively \citep{2021AJ_GBrandt2021}. This cannot be ruled out by the dynamical masses. HR~8799~e and $\beta$~Pic~c have the same $K$-band absolute magnitude within the errors, $12.94\pm0.02$\,mag and $12.9\pm0.1$\,mag, respectively, and their masses are also consistent at 0.6$\sigma$. 

The above discussion assumes that hot-start models are appropriate for deriving a substellar cooling age. If instead there was significant entropy loss in the formation of HR~8799~e, then it would be younger. Perhaps the initial entropy could even be tuned to match the age of the BPMG in a warm-start scenario. A younger age could also compensate for higher masses when considering the system's long-term stability.

\section{Conclusions}\label{sec:conclusions}

In this letter, we determine a dynamical mass for HR~8799~e of \fiducialmass by assuming that planets c, d and e share the same mass to within $\approx$20\%. Marginalizing over a larger range of mass ratios for all four planets yields a dynamical mass of \worstcasemass for HR~8799~e. We favor the more precise mass for HR~8799~e given that the planets' similar spectra and luminosities strongly suggest similar masses. 

Our dynamical mass for HR~8799~e is 2\,$\Mjup$ (1.2$\sigma$) higher than previous estimates based on hot-start models (e.g., $7.2 \pmoffs{0.6}{0.7}\,\Mjup$; \citealp{2018AJ....156..192W}). 
We rule out, with 99.7\% confidence, any planets with masses greater than $\approx6 \Mjup$ and semi-major axes between $\approx$3 au\,and $\approx 8$\,au, as well as any additional $7\,\Mjup$ or larger planets between 8 and 16\,au.

We compute an updated bolometric luminosity for HR~8799~e and use hot-start evolutionary models to derive a substellar cooling age. We find $42\pmoffs{24}{16}$\,Myr if we exclude the high-mass ($>$12\,\Mjup) portion of our mass posterior, based on the low luminosity of HR~8799~e and the stability analysis of \citet{2018AJ....156..192W}. This is consistent with both the Columba association and $\beta$~Pictoris moving group. Notably, the masses and absolute magnitudes of HR~8799~e and $\beta$~Pic~c are consistent within $<$1$\sigma$.

HR~8799~e, as the innermost planet on a $\approx$50-year period, induces about 75\% of the proper motion anomaly over the $\approx$25-year \hipparcos-\gaia baseline. The uncertainty in our dynamical mass is dominated by the \gaia proper motion precision of HR~8799. Improved astrometric precision in future \gaia data releases will translate directly to improved mass measurements for the HR~8799 planets, especially for HR~8799~e and d.

\software{astropy \citep{astropy:2013, astropy:2018},
          scipy \citep{2020SciPy-NMeth},
          numpy \citep{numpy1, numpy2},
          \htofcodename \citep{htof_zenodo, MirekHTOFtemporary},
          REBOUND \citep{rebound_2012_main},
          Jupyter (\url{https://jupyter.org/}).
          }

\acknowledgments
{G.~M.~B. is supported by the National Science Foundation (NSF) Graduate Research Fellowship under grant no. 1650114.

The orbits of the HR~8799 planets used in this work can be found at \url{whereistheplanet.com}. We thank Jason Wang, Matas Kulikauskas, and Sarah Blunt for building and hosting this wonderful web service, and for making their orbital posteriors open and freely available. We thank the anonymous referee for helpful comments that improved the quality of this work.

This work has made use of data from the European Space Agency (ESA) mission {\it Gaia} (\url{https://www.cosmos.esa.int/Gaia}), processed by the {\it Gaia} Data Processing and Analysis Consortium (DPAC, \url{https://www.cosmos.esa.int/web/Gaia/dpac/consortium}). Funding for the DPAC has been provided by national institutions, in particular the institutions participating in the {\it Gaia} Multilateral Agreement.

G.-D.\ M.\ acknowledges the support of the German Science Foundation (DFG) priority program SPP~1992 ``Exploring the Diversity of Extrasolar Planets'' (MA~9185/1-1).
G.-D.\ M.\ acknowledges support from the Swiss National Science Foundation under grant BSSGI0\_155816 ``PlanetsInTime''.
Parts of this work have been carried out within the framework of
the NCCR PlanetS supported by the Swiss National Science Foundation.

This work made use of the \htofcodename code. We used version \htofversion \citep{htof_zenodo}.}

\bibliographystyle{aasjournal}
\bibliography{refs.bib}

\end{document}